\documentclass[11pt, oneside]{article}

\usepackage{jheppub}

\usepackage{amsmath}
\usepackage{amssymb}
\usepackage{amsfonts}
\usepackage{amsthm}
\usepackage{amsbsy}
\usepackage{array}
\usepackage{multicol}
\usepackage{pgfkeys}
\usepackage{mathtools}
\usepackage{braket}
\usepackage{subcaption}
\usepackage{dsdshorthand}
\usepackage[dvipsnames]{xcolor}
\usepackage{youngtab}
\usepackage{graphicx}
\usepackage{makecell}
\usepackage{notoccite}

\newcommand \AdS {\mathrm{AdS}}
\newcommand \arcsinh {\mathop{\mathrm{arcsinh}}}

\title{
Monodromy Pinning Defects in the Critical $\mathrm{O}(2N)$ Model
}
\author{Petr Kravchuk, Alex Radcliffe}
\affiliation{
Department of Mathematics, King's College London, Strand, London, WC2R 2LS, UK
}
\date{}
\abstract{
We investigate a novel class of defects in the critical $\mathrm{O}(2N)$ model that preserve conformal symmetry along the defect, but not the symmetry under rotations transverse to the defect. Instead, they only preserve a combination of transverse rotations and a global symmetry. These defects are constructed as IR fixed points of RG flows originating at monodromy defects, triggered by a relevant operator with non-zero transverse spin. Using large-$N$ and $4-\varepsilon$ expansions, we compute leading-order scaling dimensions of defect operators and the one-point functions of the bulk fields. In various limits this theory coincides with the monodromy defect or the pinning field defect, and we compare our results to existing results for these defects. 
}

\begin{document}

\maketitle
\newpage
\section{Introduction}

A defect conformal field theory (DCFT) must be symmetric under conformal transformations along the defect, and will also typically be symmetric under rotations transverse to the defect.  It is however possible to have a DCFT that is not symmetric under transverse rotations, which we shall call a ``spinning DCFT''. Some supersymmetric examples of spinning DCFTs are known such as Gukov-Witten defects in $\mathcal{N}=4$ Super Yang-Mills~\cite{Agmon:2020pde}, and vortex loops in ABJM theory~\cite{Drukker:2008jm}.\footnote{To the best of the authors' knowledge, the transverse-rotation breaking properties of vortex loops in ABJM have not been made explicit in any published work. It will however be discussed in the upcoming work~\cite{DrukkerFuture}.} Spinning DCFTs can also arise from the fusion of non-spinning DCFTs~\cite{Kravchuk:2024qoh}. Non-supersymmetric spinning DCFTs have not been widely studied. In this paper, we discuss a simple example of such a defect in the $\mathrm{O}(2N)$ model. In our spinning DCFT the breaking of transverse rotations is mild in the sense that a combination of transverse rotations and a global symmetry is preserved; this is similar to the other examples mentioned above.

To describe these defects, we consider the critical $\mathrm{O}(2N)$ model in Euclidean $d$-dimensional space, which has the action
\begin{align}
	S=\int d^dx \p{\frac{1}{2}\partial_\mu\phi^a(x)\partial^\mu\phi^a(x)+\frac{\lambda}{4}\p{\phi^a(x)\phi^a(x)}^2}\label{eq:action1},
\end{align}
where $\phi^a$ for $a=1\dots 2N$ are real scalar fields, and $\lambda$ has been tuned to criticality.\footnote{We are working in a regularization scheme where at criticality there is no bare mass term in the action.} 

We first introduce a monodromy defect~\cite{Giombi:2021uae, Ghosh:2021ruh, Gimenez-Grau:2022czc, Barrat:2022psm, Liendo:2019jpu, Billo:2013jda, Yamaguchi:2016pbj, Soderberg:2017oaa, Gaiotto:2013nva, Metlitski:2008svp} along a codimension-2 hyperplane, which we will later perturb by a relevant operator. To this end, we group pairs of real fields into complex fields, $\Phi^I=\phi^{2I-1}+i\phi^{2I}$ for $I=1\dots N$. We shall use $\vec{y}\in \R^{d-2}$ as a coordinate on the defect, and $r\in[0, \infty)$, $\theta\in\R/(2\pi\Z)$ as polar coordinates orthogonal to the defect. The monodromy defect is produced by the requirement that $\Phi^I$ has the monodromy,
\begin{align}
	\Phi^I(\vec{y}, r, \theta+2\pi)=e^{2\pi i v}\Phi^I(\vec{y}, r, \theta),\label{eq:monodromy1}
\end{align}
where $v\in\R$ is a parameter. Without loss of generality, we shall choose $v\in [0, 1)$.

\begin{figure}[t]
	\centering
	\includegraphics[width=0.48\linewidth]{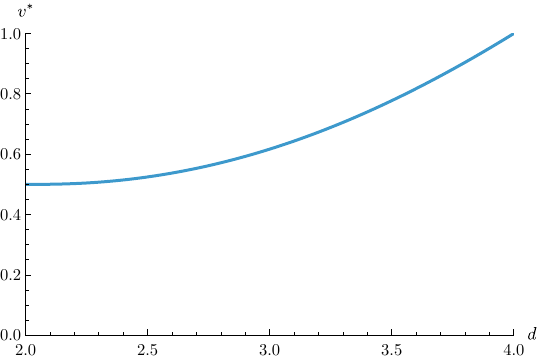}
	~
	\raisebox{15pt}{\includegraphics[width=0.48\linewidth]{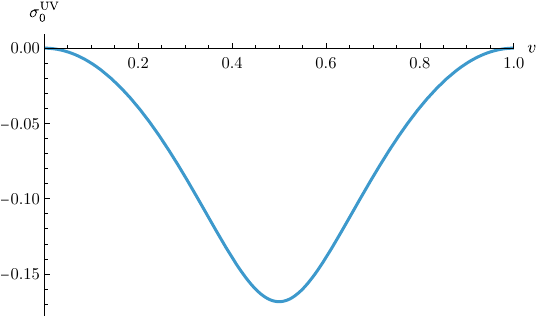}}
	\caption{Left: a plot of $v^*$ at large-$N$ for $2<d<4$. Right: a plot of $\sigma_0^{\mathrm{UV}}$ in $d=3$ in the monodromy DCFT for $0\leq v\leq 1$.}
	\label{fig:vStar}	\label{fig:sigma}
\end{figure}

The $\mathrm{O}(2N)$ CFT with a monodromy defect has been studied in~\cite{Giombi:2021uae, Ghosh:2021ruh, Gimenez-Grau:2022czc, Barrat:2022psm, Liendo:2019jpu, Billo:2013jda, Yamaguchi:2016pbj, Soderberg:2017oaa, Gaiotto:2013nva}. In this model, the bulk-to-defect expansion of $\Phi^I$ contains defect operators $\hat\Psi^I_s$ with transverse spin $s$ constrained by $s\in\Z+v$. At large $N$, the scaling dimensions of these defect operators are given by\footnote{There is an alternate set of boundary conditions, where we let $\hat\Delta_s=\frac{d-2}{2}-\sqrt{\sigma_0^{\mathrm{UV}}+s^2}$, but throughout this work,~\eqref{eq:dimensions} is the choice we take. More details about this are given in~\cite{Giombi:2021uae}.}
\begin{align}
	\hat\Delta_s=\frac{d-2}{2}+\sqrt{\sigma_0^{\mathrm{UV}}+s^2},\quad (s\in v+\Z)
	\label{eq:dimensions}
\end{align}
where $\sigma_0^{\mathrm{UV}}$ is a function of $v$ and $d$ but is independent of $s$. A plot of $\sigma_0^{\mathrm{UV}}$ as a function of $v$ in $d=3$ is shown in figure~\ref{fig:sigma}.

\begin{figure}[t]
	\centering
	\includegraphics[width=0.7\linewidth]{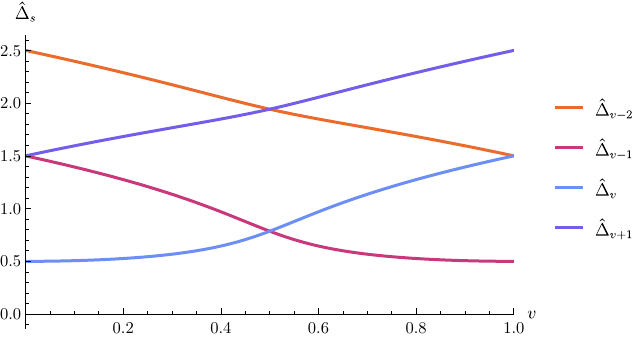}
	\caption{A plot of the scaling dimension of $\hat{\Psi}_{v+\ell}$ in $d=3$ in the monodromy DCFT for $0\leq v\leq 1$ for $\ell\in\{-2, -1, 0, 1\}$. Vertical dashed lines indicate positions of $v^*\approx 0.617$ and $1-v^*\approx 0.383$, while the horizontal dashed line indicates the marginal dimension $d-2$.}
	\label{fig:UVDimensions.pdf}
\end{figure}

We are interested in those operators $\hat\Psi^I_s$ that are relevant, i.e.\ $\hat\De_s<d-2$. For $2<d<4$ and to leading order in large $N$ the relevant operators can be identified as follows. Let $v^*$ be the value of $v$ for which $\hat\Delta_{v}=d-2$; it can be checked that there is a unique such value in $[0,1)$. Then, for any $v$, $\hat{\Psi}^I_s$ is a relevant defect operator if and only if $|s|<v^*$; this can happen only for $s\in \{v,v-1\}$. In 3d, the large-$N$ value of $v^*$ is given by
\begin{align}
	v^*_{3d}\approx 0.616797.
\end{align}
For other spacetime dimensions the large-$N$ value of $v^*$ is plotted in figure~\ref{fig:vStar}. In figure~\ref{fig:UVDimensions.pdf} we plot the scaling dimensions of the leading operators $\hat{\Psi}_{s}$ in $d=3$, as a function of $v$. We see that indeed only $\hat{\Psi}_{v}$ and $\hat{\Psi}_{v-1}$ ever become relevant.

In this work, we shall look at the RG flow triggered by such relevant operators, and in particular at the DCFT at the IR fixed point. Since $\hat \Psi_s$ has non-zero transverse spin $s$, the resulting DCFT may and in fact generally turns out to be a spinning DCFT. In the case where $v=\half$, the possibility of studying a spinning DCFT in this way was previously raised \cite{Billo:2013jda}. We shall call the IR DCFT a ``monodromy pinning DCFT,'' as in $v=0$, $d=3$ the IR fixed point is given by the pinning field defect in the $\mathrm{O}(2N)$ model, as examined in~\cite{Cuomo:2021kfm}. It is worth noting that the rotational symmetry is not completely broken, but rather a specific combination of a rotation and an internal symmetry is preserved. This will be discussed in section~\ref{sec:RG}. The various continuous symmetries of each of these models are summarized in table~\ref{table:symmetries}.

\begin{table}[h]
	\centering
	\begin{tabular}{|l|c|c|c|}
		\hline
		& {Monodromy Defect} &\makecell{{Monodromy Pinning}\\{Defect}} & \makecell{{Pinning Field}\\{Line Defect}} \\
		\hline
		\makecell{{Spacetime}\\{Symmetry}} & $\mathrm{SO}_0(d-1, 1)\times \mathrm{SO}(2)$ & $\mathrm{SO}_0(d-1, 1)$ & $\mathrm{SL}_2(\mathbb{R})\times \mathrm{SO}(d-1)$ \\
		\hline
		\makecell{{Internal}\\{Symmetry}} & $\mathrm{U}(N)$ & $\mathrm{U}(N-1)$ & $\mathrm{O}(2N)$ \\
		\hline
		\makecell{{Mixed}\\{Symmetry}} & -- & $\mathbb{R}$ or $\mathrm{U}(1)$ & -- \\
		\hline
	\end{tabular}
	\caption{\label{table:symmetries}The continuous symmetries of monodromy defects, monodromy pinning defects, and pinning field defects, assuming that $v\neq0, \tfrac{1}{2}$.}
\end{table}

We shall propose a description for monodromy pinning DCFTs at large $N$ using techniques inspired by~\cite{Cuomo:2021kfm}. At large $N$, we shall calculate the scaling dimensions of various defect operators in this theory and a bulk one-point function. We also verify the presence of the displacement operator as well as a tilt operator corresponding to the breaking of transverse rotations. 

\begin{figure}[t]
	\centering
	\includegraphics[width=0.5\linewidth]{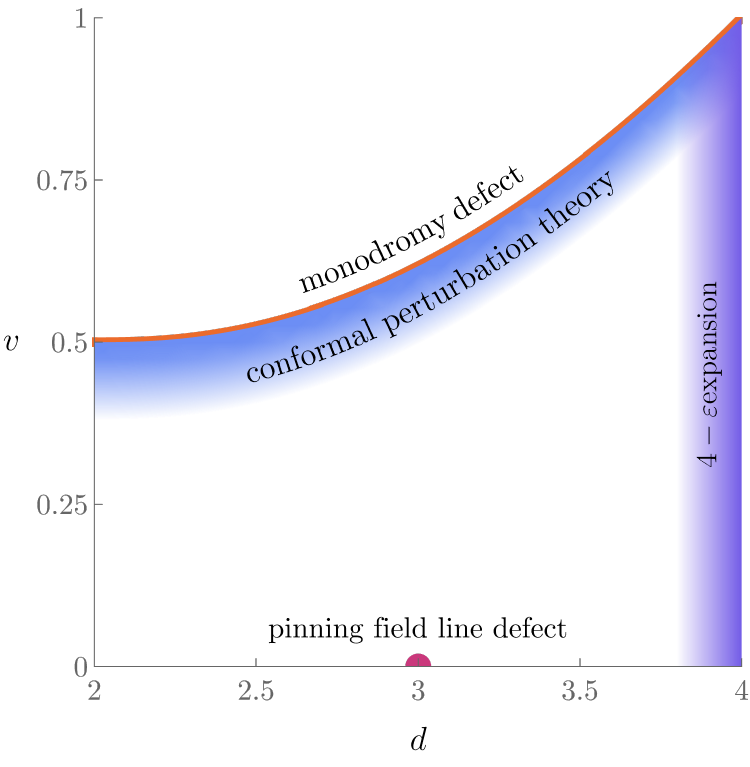}
	\caption{The relationship between various limits of the monodromy pinning DCFT in the $\mathrm{O}(2N)$ model. The monodromy pinning DCFT exists at large $N$ everywhere below the monodromy defect line.}
	\label{fig:schematic}
\end{figure}

In various limits, this DCFT coincides with theories which can be examined via other means. These relationships are illustrated in figure~\ref{fig:schematic}. For example, in $d=3$ and $v=0$, the monodromy pinning DCFT coincides with the pinning field defect studied in~\cite{Cuomo:2021kfm}; in the limit $v\nearrow v^*$, the theory approaches the monodromy defect theory~\cite{Giombi:2021uae, Ghosh:2021ruh, Gimenez-Grau:2022czc, Barrat:2022psm, Liendo:2019jpu, Billo:2013jda, Yamaguchi:2016pbj, Soderberg:2017oaa, Gaiotto:2013nva} and can be examined through conformal perturbation theory; and for $d=4-\varepsilon$, this theory can be studied in an $\varepsilon$-expansion. We shall compare the results obtained through these methods to verify that we have identified the correct IR fixed point.

\textbf{Structure of the paper:} In section~\ref{sec:rg_flows} we discuss relevant defect operators in the $\mathrm{O}(2N)$ model with a monodromy defect, and look at the symmetries of the RG flows triggered by these operators. In section~\ref{sec:large_n} we describe the model at large $N$, and calculate the scaling dimensions of defect operators, and a bulk one-point function. In section~\ref{sec:CPT}, we perform a basic analysis of the limit $v\nearrow v^*$ in conformal perturbation theory. In section~\ref{sec:epsilon}, we study the monodromy pinning defect in $d=4-\varepsilon$ to leading order in $\varepsilon$, and compare this to the results in the large-$N$ expansion. We conclude in section~\ref{sec:conclusions}. The appendices clarify various technical details.

\section{Defect RG flows from a monodromy defect in the $\mathrm{O}(2N)$ model}
\label{sec:rg_flows}
\subsection{Symmetries of monodromy defects in the $\mathrm{O}(2N)$ model}
\label{sec:symmetries}

Generally, adding a monodromy defect into the $\mathrm{O}(2N)$ model means requiring that
\begin{align}
	\phi^a(\vec{y}, r, \theta+2\pi)=G^{a}{}_{b}\phi^b(\vec{y}, r, \theta),\label{eq:monodromy2}
\end{align}
for some $G\in\mathrm{O}(2N)$. In this paper we only consider the special case~\cite{Giombi:2021uae} where
\begin{align}
	G=\begin{pmatrix}
		\cos(2\pi v)&-\sin(2\pi v)&0&0&\dots&0&0\\
		\sin(2\pi v)&\cos(2\pi v)&0&0&\dots&0&0\\
		0&0&\cos(2\pi v)&-\sin(2\pi v)&\dots&0&0\\
		0&0&\sin(2\pi v)&\cos(2\pi v)&\dots&0&0\\
		\vdots&\vdots&\vdots&\vdots&\ddots&\vdots&\vdots\\
		0&0&0&0&\dots&\cos(2\pi v)&-\sin(2\pi v)\\
		0&0&0&0&\dots&\sin(2\pi v)&\cos(2\pi v)\\
	\end{pmatrix},
\end{align}
which gives us the monodromy in~\eqref{eq:monodromy1} in terms of the complex fields $\Phi^I=\phi^{2I-1}+i\phi^{2I}$.

This leads to $\Phi^I$ not being single-valued (unless $v=0$). One way around this is to view the defect $\mathcal{D}$ as the boundary of a codimension-1 submanifold $\Sigma\subset\R^d$---for example, of the hypersurface at $\theta=0$. We can then introduce the monodromy around $\mathcal{D}$ by letting $\Phi^I$ have a discontinuity at $\Sigma$. $\Sigma$ is clearly topological as the action is invariant under $\phi^a\mapsto G^{a}{}_b\phi^b$. It is essentially the topological defect associated to the global symmetry transformation $G\in \mathrm{O}(2N)$. 

Inserting a (non-topological) flat codimension-$q$ defect into a $d$-dimensional CFT necessarily breaks the conformal symmetry down to at most the subgroup of transformations that preserve the location of the defect. These transformations are generated by transverse rotations around the defect, and conformal transformations along the defect, giving the group $\mathrm{SO}_0(d-q+1, 1)\times\mathrm{SO}(q)$. In the case of a monodromy defect, $q=2$, and the entire $\mathrm{SO}_0(d-1, 1)\times \mathrm{SO}(2)$ group of symmetries is preserved in the monodromy defect DCFT with action given by~\eqref{eq:action1}.\footnote{There is a small caveat that transverse rotations move $\Sigma$. However, since $\Sigma$ is topological, at least for infinitesimal rotations it can be moved back after the rotation.\label{footnote:footnote}}

One consequence of the insertion of a defect is that as we break the symmetry under translations normal to the defect, not all of the components of the stress tensor are conserved. Specifically, we expect that
\begin{align}
	\partial_\mu T^{\mu i}(\vec{y}, r, \theta)=D^i(\vec{y})\delta^{(2)}(r, \theta),
\end{align}
for a defect operator $D^i(\vec{y})$ where the index $i$ runs over the directions perpendicular to the defect, and $\delta^{(2)}(r, \theta)$ is a delta function in the plane normal to the defect. $D^i$ is called the displacement operator~\cite{Billo:2016cpy}. As the scaling dimension of the stress tensor is protected, the scaling dimension of $D^i$ must be $d-1$, and its $\mathrm{SO}(2)$-spin transverse to the defect must be $1$.

The subgroup of internal symmetries that are preserved under the insertion of the monodromy defect is given by the centralizer of $G$ in $\mathrm{O}(2N)$~\cite{Giombi:2021uae},
\begin{align}
	C_{\mathrm{O}(2N)}(G)&=\begin{cases}
		\mathrm{O}(2N)&\text{if }v\in\{0, \tfrac{1}{2}\}\\
		\mathrm{U}(N)&\text{otherwise,}\\
	\end{cases}\label{eq:centralizer}
\end{align}
where $\mathrm{U}(N)$ is a subgroup of $\mathrm{O}(2N)$ acting on the complex fields $\Phi^I$ in the fundamental representation.

In the bulk, regardless of the value of $v$, we retain a local $\mathrm{O}(2N)$ internal symmetry, and we have $N(2N-1)$ real Noether currents associated with this, given by
\begin{align}
	J^{ab}_\mu=\phi^a\partial_\mu\phi^b-\phi^b\partial_\mu\phi^a.\label{eq:current}
\end{align}
Their dimension is protected, $\Delta_J=d-1$. For $v\neq0, \tfrac{1}{2}$, some of these currents are discontinuous across $\Sigma$, but the currents corresponding to the $\mathrm{U}(N)$ symmetry are continuous across $\Sigma$. These $\mathrm{U}(N)$ currents are given by the skew-Hermitian matrix
\begin{align}
	\mathcal{J}^{IJ}_\mu&=iJ^{2I-1, 2J-1}-J^{2I-1, 2J}+J^{2I, 2J-1}-iJ^{2I, 2J}\nn\\
	&=i(\bar\Phi{}^I\partial_\mu\Phi^J-\bar\Phi{}^J\partial_\mu\Phi^I).
\end{align}
As was mentioned in~\eqref{eq:centralizer}, the $\mathrm{O}(2N)$ internal symmetry group is broken to $\mathrm{U}(N)$ if $v\neq0, \tfrac{1}{2}$. This symmetry breaking will, however, not give a tilt operator\footnote{Tilt operators generally arise when a defect breaks a part of continuous bulk symmetries. We discuss tilt operators in more detail in section~\ref{sec:RG}.} on $\mathcal{D}$, as the broken currents are not continuous across $\Sigma$ and so their divergences are not localized on $\cD$. Another way to see this is that tilt operators can be added to the action with position-dependent couplings, and therefore reflect our ability to vary the symmetry-breaking parameters along $\cD$. However, in the present case the symmetry-breaking parameter $G$ cannot be varied along $\cD$ as it has to be constant on $\Sigma$.

Finally, we note that there exists a discrete symmetry under the reflection along the defect,
\begin{align}
	R^{\parallel}: (y_1, y_2, \dots y_{d-2})\mapsto (-y_1, y_2, \dots y_{d-2}).
\end{align}
If $v\in\{0, \tfrac{1}{2}\}$ then we also have a discrete symmetry under the reflection transverse to the defect,
\begin{align}
	R^\perp:\theta\mapsto2\pi-\theta.
\end{align}

\subsection{RG flow triggered by $\Re(\hat{\Psi}^1_s)$ for $v\not\in\{0, \tfrac{1}{2}\}$}
\label{sec:RG}

Throughout this section, we shall assume that $v\not\in\{0, \tfrac{1}{2}\}$, and we shall address the remaining special cases in section~\ref{sec:v}. As was mentioned in the introduction, the defect operators $\hat{\Psi}_s$ have transverse spin $s\in\mathbb{Z}+v$. It is known that $\hat{\Psi}_s$ is relevant at large $N$ if and only if $|s|\leq v^*$, where $v^*$ is a function of $d$. We hence study the relevant perturbation by
\begin{align}
	\int_{\mathcal{D}}d^{d-2}\vec{y}\,
	\Re(h^I\hat{\Psi}^I_s(\vec{y}))
\end{align}
for some complex coupling $h^I$ and $s=v$ or $s=v-1$.  Due to the $\mathrm{U}(N)$ internal symmetry of the monodromy defect DCFT, we can assume without loss of generality that $h^I=h\delta^{I1}$ where $h\in\R$.

It is worth noting that if $0<v,1-v<v^*$, then both $\hat\Psi_v$ and $\hat \Psi_{v-1}$ are relevant. It is hence possible to study a perturbation by either of these operators, or by a combination of them both. As we will see below, perturbation by these operators preserve different symmetries, and it therefore makes sense to study the perturbation by just one of them. We shall hence leave the study of a simultaneous perturbation by $\hat\Psi_v$ and $\hat \Psi_{v-1}$ for future work.

If we interchange $\Phi^I$ and $\bar\Phi{}^I$, then the monodromy defect with monodromy $v$ is mapped to the monodromy defect with monodromy $1-v$. In terms of defect operators, this interchanges $\Re(\hat{\Psi}_v)$ and $\Re(\hat{\Psi}_{v-1})$, so we shall without loss of generality assume that we are adding $\Re(\hat{\Psi}_v)$. This leads us to consider the action,
\begin{align}
	S=\int d^dx \p{\frac{1}{2}\partial_\mu\bar\Phi{}^I(x)\partial^\mu\Phi^I(x)+\frac{\lambda}{4}\p{\bar\Phi{}^I(x)\Phi^I(x)}^2}+h\int_{\mathcal{D}}d^{d-2}\vec{y}\,
	\Re(\hat{\Psi}^1_v(\vec{y})).\label{eq:action3}
\end{align}

We can see that in the action~\eqref{eq:action3}, the $\mathrm{U}(N)$ internal symmetry is broken to $\mathrm{U}(N-1)$, and the symmetry under transverse rotations is also broken --- $\hat{\Psi}^1_v(\vec{y})$ has transverse spin $v$. There is however also a combination of a broken $\mathrm{U}(N)$ generator and a transverse rotation that is preserved. Specifically, we shall let $Q$ be the internal symmetry charge that acts on $\Phi^I$ as $[Q, \Phi^I]=i\delta^{I1}\Phi^1$. We then let $J$ be the charge associated with rotations that acts as $[J, \Phi^I]=i\partial_\theta\Phi^I$. If we now let 
\begin{align}
	U_s=sQ-J,\label{eq:charge}
\end{align}
then $[U_s, \hat{\Psi}^1_s]=0$, and $U_v$ generates a (generally non-compact) $\mathfrak{u}(1)$-symmetry of the theory that is preserved by the RG flow. The symmetries of the IR and UV theories (and of pinning field line defects) are given in table~\ref{table:symmetries}. 

Just as breaking the symmetry under translations perpendicular to the defect gives us a displacement operator, breaking an internal symmetry gives rise to defect operators called tilt operators. The Ward identities associated to the broken generators are modified to
\begin{align}
	\partial^\mu \mathcal{J}^{1I}_\mu(x)=t^I(\vec{y})\delta^{(2)}(r, \theta),
\end{align}
where $t^1$ is a real tilt operator transforming in the trivial representation of $\mathrm{U}(N-1)$, and $t^{\hat{I}}$ for $\hat{I}=2\dots N$ is a complex tilt operator transforming in the fundamental representation of $\mathrm{U}(N-1)$. As the symmetry under transverse rotations is also broken, its Ward identity is also modified, and there is a corresponding tilt operator. However, as the symmetry generated by $U_v$ is preserved, this tilt operator must be proportional to $t^1$. The scaling dimension of these tilt operators is given by
\begin{align}
	\Delta_{t}=d-2,
\end{align}
as the dimension of the current is protected.

We expect that in the IR the $\mathrm{SO}(d-2+1, 1)$ conformal symmetry along the defect is restored. The discrete symmetry $R^{\parallel}$ is preserved along the RG flow.

\subsection{$v=0$ and $v=\tfrac{1}{2}$}
\label{sec:v}

In the cases where $v=0$ or $v=\tfrac{1}{2}$, the monodromy can be written in terms of the real fields as
\begin{align}
	\phi^a(\vec{y}, r, \theta)=\pm\phi^a(\vec{y}, r, \theta+2\pi),
\end{align}
where the positive sign is taken if $v=0$, and the negative sign is taken if $v=\tfrac{1}{2}$. This allows us to generalize to the cases where we have an odd number of real fields, which we shall represent by taking $N\in\tfrac{1}{2}\Z$. As the defect now preserves the full $\mathrm{O}(2N)$ symmetry, the defect operators $\hat \Psi^I_s$ can be reorganized into vector multiplets of $\mathrm{O}(2N)$. We denote such operators by $\hat \chi^a_s$. In general, these are not Hermitian as they still carry a definite transverse spin $s$, $\bar{\hat \chi^a_s} = \hat \chi^a_{-s}$. See appendix~\ref{app:OPE} for details.

In the case $v=0$ we have $\hat \Psi^1_0=\hat\chi^1_0+i\hat\chi^2_0$ and our perturbation is by $\Re(\hat \Psi^1_0)=\hat \chi^1_0$, which preserves $\mathrm{O}(2N-1)$ symmetry and transverse rotations. In $d=3$ the resulting IR DCFT coincides with the pinning field line defect, see~\cite{Cuomo:2021kfm} and references therein. For other values of $d$ the $v=0$ defect can be seen as a ``transdimensional''~\cite{deSabbata:2024xwn} version of the pinning field defect.

In the case $v=\half$ we still have $\hat \Psi^1_\half=\hat\chi^1_\half+i\hat\chi^2_\half$. However, now the operators $\hat\chi^1_\half$ and $\hat\chi^2_\half$ are not Hermitian and both survive after taking $\Re(\hat\Psi^1_\half)$. Therefore, our perturbation is a combination of $\hat\chi^{1,2}_{\pm\half}$ and preserves only an $\mathrm{O}(2N-2)$ subgroup of $\mathrm{O}(2N)$. The particular linear combination of these operators is protected by CPT and the mixed charge $U_\half$. Note that $\dim \mathrm{O}(2N)-\dim \mathrm{O}(2N-2)=4N-3$, so we will have to identify $2N-2$ additional tilt operators for $v=\half$ in addition to the $2N-1$ tilt operators present for generic $v$.

For $v=\half$ we could instead consider the RG flow triggered by $\Re(\chi^1_v)$, which would preserve a larger subgroup of internal symmetries $\mathrm{O}(2N-1)$. However, this comes at the cost of losing the symmetry generated by $U_v$, and we shall not consider this case in this paper.

\section{Monodromy pinning DCFTs at large $N$}
\label{sec:large_n}

In this section, we shall outline the description of the monodromy pinning DCFTs at large $N$. Specifically, we shall begin by performing a Weyl transformation into $\AdS_{d-1}\times S^1$. This will map the defect to the boundary of $\AdS_{d-1}$. We shall then perform the Hubbard-Stratonovich transformation and argue that the IR fixed point of~\eqref{eq:action3} can be described by a suitable boundary condition for $\AdS_{d-1}$, following~\cite{Cuomo:2021kfm, Giombi:2021uae}. We shall then calculate various observables in this theory, such as scaling dimensions of defect operators and the one-point function of $\Phi^I(x)$. We shall compare these results to the results of~\cite{Cuomo:2021kfm} and to the results of conformal perturbation at $v=0$ and $v=v^*$, respectively. This will help to confirm that we have identified the correct IR fixed point.

\subsection{Weyl transformation to $\AdS_{d-1}\times S^1$ description}

In order to expand the critical $\mathrm{O}(2N)$ model at large $N$, we begin with the action
\begin{align}
	S=\int d^dx \p{\frac{1}{2}\partial_\mu\bar\Phi{}^I(x)\partial^\mu\Phi^I(x)+\frac{\lambda}{N}\p{\bar\Phi{}^I(x)\Phi^I(x)}^2}\label{eq:LargeNAction}.
\end{align}
We have rescaled the coupling $\lambda$ as compared with~\eqref{eq:action1}, so as to obtain a smooth large-$N$ limit with $\lambda=\mathcal{O}(1)$. Our method for studying the pinning monodromy DCFT at large $N$ is inspired by the treatment of the pinning field defect in~\cite{Cuomo:2021kfm} and of the monodromy defect in~\cite{Giombi:2021uae}. We hence begin by performing the Weyl transformation,
\begin{align}
	ds_{\R^d}^2=d\vec{y}^2+dr^2+r^2d\theta^2\longmapsto \frac{1}{r^2}ds_{\R^d}^2=\frac{d\vec{y}^2+dr^2}{r^2}+d\theta^2.
\end{align}
The result is the metric for $\AdS_{d-1}\times S^1$ (with the AdS radius equal to 1), with $r$ and $y$ now acting as Poincar\'e coordinates on $\AdS_{d-1}$, and $\theta$ as a coordinate on $S^1$. We shall use $X$ as a shorthand for the AdS coordinates $(r, \vec{y})$, and $x$ as a shorthand for the complete $\AdS_{d-1}\times S^1$ coordinates $(r, \vec{y}, \theta)$. We shall also use $g$ to denote the metric on $\AdS_{d-1}\times S^1$, and $G$ for the metric on $\AdS_{d-1}$. This choice of Weyl frame is convenient because conformal transformations along the defect are realized as AdS isometries.

We then begin with the critical $\mathrm{O}(2N)$ action in $\AdS_{d-1}\times S^1$,
\begin{align}
	S_{\AdS_{d-1}\times S^1}=\int d^dx \sqrt{g} \p{\frac{1}{2}\nabla_\mu\bar\Phi{}^I\nabla^\mu\Phi^I-\frac{(d-2)^2}{8}\bar\Phi{}^I\Phi^I+\frac{\lambda}{N}\p{\bar\Phi{}^I\Phi^I}^2},\label{eq:AdSAction}
\end{align}
where we have added a conformal mass term for $\AdS_{d-1}\times S^1$ determined by its Ricci scalar, $\mathcal{R}=-(d-2)(d-1)$, and $g$ is the metric for $\AdS_{d-1}\times S^1$. We perform a Hubbard-Stratonovich transformation to obtain
\begin{align}\label{eq:SHS}
	S_{\text{HS}}&=\int d^dx \sqrt{g} \p{\frac{1}{2}\nabla_\mu\bar\Phi{}^I\nabla^\mu\Phi^I-\frac{(d-2)^2}{8}\bar\Phi{}^I\Phi^I+\frac{1}{2}\sigma\bar\Phi{}^I\Phi^I}\nn\\
	&=\frac{1}{2}\int d^dx \sqrt{g} \p{\bar\Phi{}^I\p{-\nabla^2-\frac{(d-2)^2}{4}+\sigma}\Phi^I},
\end{align}
such that a large-$N$ expansion of the theory corresponds to a saddle-point expansion of this theory in $\sigma$.\footnote{A Hubbard-Stratonovich transformation generally also involves an additional term in the Lagrangian, $\tfrac{\sigma(x)^2}{4\lambda},$ so that the equation of motion for $\sigma$ is $\sigma=\lambda\Phi^I\bar\Phi{}^I$. It can however be shown that this term can be dropped without affecting the IR fixed point~\cite{Giombi:2016ejx}.} The monodromy of the bulk fields is given by
\begin{align}
	\Phi^I(r, \vec{y}, \theta+2\pi)&=e^{2\pi i v}\Phi^I(r, \vec{y}, \theta),
\end{align}
however
\begin{align}
	\sigma(r, \vec{y}, \theta+2\pi)&=\sigma(r, \vec{y}, \theta);
\end{align}
in other words, $\sigma$ has no monodromy. 

Since we expect to obtain a conformal defect in the IR, the saddle point value $\s_0^\mathrm{IR}$ of $\s$ should be a constant. Therefore, to the leading order in $N$, the field $\Phi$ behaves as a free field with a mass determined by $\s_0^\mathrm{IR}$ through~\eqref{eq:SHS}.

To describe the pinning monodromy DCFT in this picture it is helpful to perform the Fourier decomposition 
\begin{align}
	\Phi^I(x)=\sum_{s\in\mathbb{Z}+v}e^{is\theta}\Phi^I_s(X),\qquad\bar\Phi{}^I(x)=\sum_{s\in\mathbb{Z}+v}e^{-is\theta}\bar\Phi{}^I_s(X)
\end{align}
Note that the modes $\Phi^I_s$ are simply Fourier/Kaluza-Klein modes of $\Phi^I$, and are therefore distinct from $\hat{\Psi}^I_s$, which are defect primaries living at the boundary of $\AdS_{d-1}$.

Adding the operator $\Re(\hat{\Psi}^1_v)$ to the defect action can be seen as adding a source term $J$ for the boundary value of the component $\Re(\Phi^1_s)$. By the GKPW AdS/CFT dictionary~\cite{Witten:1998qj,Gubser:1998bc}, this imposes a boundary condition on the non-normalizable mode of $\Re(\Phi^1_v)$, which has the asymptotic form
\begin{align}
	\<\Phi^1_v(X)\>\propto r^{d-2-\Delta_v}J\label{eq:source}
\end{align}
for a scaling dimension $\De_v$ determined by the effective mass of $\Phi$ as discussed below (not to be confused with $\hat \De_v$ in~\eqref{eq:dimensions}) and a constant source $J\in \R$. This source is effectively the IR fixed-point value of the coupling $h$.

With the view towards integrating out the bulk fields $\Phi^I$, it is then convenient to define 
\begin{align}\label{eq:dePhi}
	\delta\Phi^I_s(X_1)=\Phi^I_s(X_1)-\frac{\sqrt{2 N}}{2\pi}\delta^{I1}\delta_{vs}\int d^{d-2}\vec{y}_2 \,\int d\theta_1 e^{iv\theta_1}G_{\text{b} \partial}^{v}(x_1, \vec{y}_2)J,
\end{align}
which also fixes the normalization of the source $J>0$.\footnote{Note that we insist on $J>0$ in~\eqref{eq:dePhi}. In principle, given~\eqref{eq:action3}, the phase of $J$ is determined by the RG flow. We have not studied this relationship in detail. However, we can always achieve $J>0$ by applying a transverse rotation to~\eqref{eq:action3}. We assume that this has been done.
} The defining property of $\de \Phi^I_s$ is that it has vanishing $1$-point function,
\begin{align}
	\<\de \Phi^I_s\>=0.
\end{align}
In other words, the term linear in $\de\Phi^I_s$ is cancelled in the effective action, see~\eqref{eq:deltaPhiAction} below. Above, $G_{\text{b} \partial}^v$ is a particular mode of the bulk-to-boundary propagator,
\begin{align}
	G_{\text{b}\partial}^v(x_1, \vec{y}_2)&=\lim_{r_2\to0}\frac{1}{2\pi}\int d\theta_2\, e^{-iv\theta_2}r_2^{-\Delta_v}G_{\mathrm{bb}}(x_1, x_2),
\end{align}
where the bulk-to-bulk propagator is defined as
\begin{align}
		G_{\mathrm{bb}}(x_1, x_2) &= \frac{1}{-\nabla^2+\sigma-\frac{(d-2)^2}{4}}(x_1, x_2),\label{eq:Gbb}
\end{align}
with the monodromy that
\begin{align}
	G_{\mathrm{bb}}(X_1, \theta_1, X_2, \theta_2)=e^{2\pi i v}G_{\mathrm{bb}}(X_1, \theta_1+2\pi, X_2, \theta_2)=e^{-2\pi i v}G_{\mathrm{bb}}(X_1, \theta_1, X_2, \theta_2+2\pi).
\end{align}
We also define the boundary-to-boundary propagator (which we will use later) as
\begin{align}
	G^v_{\partial\partial}(\vec{y}_1, \vec{y}_2)&=\lim_{r_1\to0}\frac{1}{2\pi}\int d\theta_1 e^{iv\theta_1}r_1^{-\Delta_v}G^v_{\mathrm{b}\partial}(x_1, \vec{y}_2).
\end{align}
Note that we use the full field $\s$ in~\eqref{eq:Gbb} instead of the saddle-point value $\s_0^\mathrm{IR}$. The reason for this will become clear in~\eqref{eq:deltaPhiAction}.

The scaling dimension $\De_v$ is determined by the effective mass of the $\Phi_v^I$ component. Recall that we use $\sigma_0^{\mathrm{IR}}$ to denote the saddle-point value of $\s$, and at the saddle-point, the bulk part of the action can be expanded in Fourier modes,
\begin{align}
	\sum_{s\in\mathbb{Z}+v}\pi\int d^{d-1}X \sqrt{G(X)} \p{\bar\Phi{}^I_s(X)\p{-\nabla^2-\tfrac{(d-2)^2}{4}+\sigma_0^{\mathrm{IR}}+s^2}\Phi^I_{s}(X)}.
\end{align}
The masses of the modes are hence given by
\begin{align}
	m_s^2=\sigma_0^{\mathrm{IR}}+s^2-\tfrac{(d-2)^2}{4},
\end{align}
and are linked to the scaling dimensions as
\begin{align}
	\Delta_s&=\frac{d-2}{2}\pm\sqrt{\frac{(d-2)^2}{4}+m_s^2}
\end{align}
We have a choice of sign in some of these modes (if allowed by unitarity), which corresponds to a choice of boundary conditions. In this work, we always impose the boundary conditions that
\begin{align}
	\Delta_s&=\frac{d-2}{2}+\sqrt{\frac{(d-2)^2}{4}+m_s^2}\\
	&=\frac{d-2}{2}+\sqrt{\sigma_0^{\mathrm{IR}}+s^2}.\label{eq:scaling}
\end{align}

We can now write the full action in terms of $\delta\Phi^I$ variables. The shift between $\delta\Phi^I_s$ and $\Phi^I_s$ eliminates the source term for $\Phi^I_s$ but produces a quadratic term in $J$, which yields
\begin{align}
	S&=\frac{1}{2}\int d^{d-1}X \sqrt{G(X)} \p{\overline{\delta\Phi}{}^I(X)\p{-\nabla^2-\tfrac{(d-2)^2}{4}+\sigma+s^2}\delta\Phi^I(X)}\nn\\
	&- N\int\,d^{d-2}\vec{y}_1\int\,d^{d-2}\vec{y}_2\, JG_{\partial\partial}^v(\vec{y}_1, \vec{y}_2) J.\label{eq:deltaPhiAction}
\end{align}
It was important for this to use the $\s$-dependent propagator in~\eqref{eq:Gbb}.

We can now integrate out the fields $\de \Phi^I$ to find the effective action for $\s$,
\begin{align}
	S_{\text{eff}}=\frac{N}{2}\Tr\log\p{-\nabla^2+\sigma-\frac{(d-2)^2}{4}}- NJ^2\int\,d^{d-2}\vec{y}_1\int\,d^{d-2}\vec{y}_2\, G^v_{\partial\partial}(\vec{y}_1, \vec{y}_2).\label{eq:action6}
\end{align}
Note that here we are taking the trace over a complex field with monodromy given as in \eqref{eq:monodromy1}.

The constant value of $J$ effectively corresponds to the fixed-point value of the $\Re(\hat{\Psi}^1_v)$ coupling and is left undetermined by the above discussion. We can find it using the following logic. Assuming that the fixed point has conformal symmetry, the one-point functions of local operators should be conformally-invariant. In the $\AdS_{d-1}\x S^1$ picture, conformal invariance becomes invariance under $\AdS$ isometries. Scalar one-point functions invariant under $\AdS$ isometries are necessarily constant. This is only consistent with~\eqref{eq:source} if
\begin{align}
	\De_v = d-2
\end{align}
which through~\eqref{eq:scaling} implies
\begin{align}
	\sigma_0^{\mathrm{IR}}=\frac{(d-2)^2}{4}-v^2.\label{eq:saddle}
\end{align}
On the other hand, $\sigma_0^{\mathrm{IR}}$ should be the saddle-point value for the action~\eqref{eq:action6} which depends on $J$; this condition fixes $J$. 

Note that $\De_v = d-2$ implies that the boundary values of the bulk fields $\Phi^I_v$ with $I\neq 1$ and $\Im(\Phi^1_v)$ become operators of dimension $d-2$, and can be identified with the tilt operators $t^{\hat I}$ and $t^1$ discussed in section~\ref{sec:RG}. For $v=\half$ this is also true for the boundary values of $\Phi_{v-1}^I$ due to $(v-1)^2=v^2$. This provides the $2N-2$ additional Hermitian tilt operators that we expect in this case, see section~\ref{sec:v}.

\begin{figure}[t]
	\centering
	\includegraphics[width=0.5\linewidth]{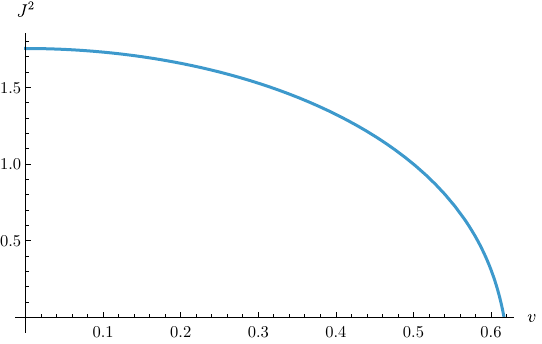}
	\caption{A plot of $J^2$ for $0\leq v\leq v^*$ in $d=3$.}
	\label{fig:JJ}
\end{figure}

The saddle-point equation implied by~\eqref{eq:action6} for a constant $\sigma_0^{\mathrm{IR}}$ is
\begin{align}
	\left.G_{\mathrm{bb}}(x_1, x_1)+J^2\left|\int d^{d-2}\vec{y}_2\, G^v_{\mathrm{b} \partial}(x_1, \vec{y}_2)\right|^2\right|_{\sigma=\sigma_0^{\mathrm{IR}}}=0.
\end{align}
The integral can be calculated using the explicit form of the bulk-to-boundary propagator at $\s=\sigma_0^{\mathrm{IR}}$,
\begin{align}
	G^v_{\mathrm{b} \partial}(x_1, \vec{y}_2)|_{\sigma=\sigma_0^{\mathrm{IR}}}&=\frac{\Gamma (d-2)e^{-iv\theta_1}}{4\pi^{\frac{d}{2}}\,\Gamma \left(\frac{d-2}{2}\right)}\left(\frac{r_1^2+(\vec{y}_1-\vec{y}_2)^2}{r_1}\right)^{2-d},
\end{align}
which yields
\begin{align}
	\int d^{d-2}\vec{y}_1 G^v_{\mathrm{b} \partial}(x_1, \vec{y}_2)|_{\sigma=\sigma_0^{\mathrm{IR}}}=\frac{e^{-iv \theta_1}}{2\pi  (d-2)},
\end{align}
and therefore
\begin{align}
	J^2=-4\pi^2(d-2)^2G_{\mathrm{bb}}(x_1, x_1).\label{eq:JJ}
\end{align}
Expressing $\sigma_0^{\mathrm{IR}}$ in terms of $v$ via~\eqref{eq:saddle}, we calculate the right-hand side explicitly in appendix~\ref{app:Gbb} as a function of $v$. We plot the resulting function in $d=3$ for $0\leq v\leq v^*$ in figure~\ref{fig:JJ}.

Knowing $J$, we can determine the one-point function of $\Phi$
\begin{align}
	\braket{\Phi^I(x)}=\de^I_1\frac{\sqrt{N}J}{2\pi(d-2)}e^{iv\theta}.
\end{align}
After Weyl transforming into flat space, this becomes
\begin{align}\label{eq:largeN1pt}
	\braket{\Phi^I(x)}_{\mathrm{flat}}=\de^I_1\frac{\sqrt{N}J}{2\pi(d-2)r^{\Delta_\Phi}}e^{iv\theta}.
\end{align}
where $\Delta_\Phi=\frac{d-2}{2}+O(N^{-1})$ is the bulk scaling dimension of $\Phi^I$ in the $\mathrm{O}(2N)$ model.

\subsection{Spectral representation of $G_{\delta\sigma\delta\sigma}$}

Define the fluctuation $\de\s$ of $\s$ by
\begin{align}
	\s = \sigma_0^{\mathrm{IR}}+ \de \s.
\end{align}
In this section we compute the bulk two-point function $\<\de \s\de\s\>$ and decompose it in the defect channel to obtain the spectrum of some $\mathrm{U}(N-1)$-neutral defect operators.

In the action, the quadratic term in $\de \s$ is given by
\begin{align}
	S_{\text{eff}}=
	N\int d^dx_1\,d^dx_2\,\sqrt{g(x_1)g(x_2)}\,\delta\sigma(x_1)G_{\delta\sigma\delta\sigma}^{-1}(x_1, x_2)\delta\sigma(x_2)
\end{align}
where,\footnote{The derivation of this is almost identical to the derivation of (4.35) in~\cite{Cuomo:2021kfm}.}
\begin{align}
	G_{\delta\sigma\delta\sigma}^{-1}(x_1, x_2)=&-\frac{1}{2}G_{\mathrm{bb}}(x_1, x_2)G_{\mathrm{bb}}(x_2, x_1)\\
	&-\frac{J^2}{8\pi^2}\left(e^{-iv\theta_{21}}G_{\mathrm{bb}}(x_1, x_2) + e^{iv\theta_{21}}G_{\mathrm{bb}}(x_2, x_1)\right),\label{eq:deltasigmadeltasigma}
\end{align}
with
\begin{align}
	\theta_{21}=\theta_2-\theta_1
\end{align}
and with the propagators on the right-hand side evaluated at $\s=\sigma_0^{\mathrm{IR}}$.

We wish to decompose $G_{\delta\sigma\delta\sigma}$ into a basis of AdS harmonic functions, and so we shall first introduce the AdS propagator,
\begin{align}
	&G_\AdS^{\Delta}(x_1, x_2)=\frac{\Gamma(\Delta)\big(2\cosh(\sigma(x_1, x_2))\big)^{-\Delta}}
	{2\pi^{\tfrac{d-2}{2}}\,\Gamma\!\left(\Delta-\frac{d}{2}+2\right)}
	{}_2F_1\!\left(\frac{\Delta}{2},\frac{\Delta+1}{2};
	\;\Delta-\frac{d}{2}+2;
	\;\frac{1}{\cosh^2(\sigma(x_1, x_2))}\right),\label{eq:propagator}
\end{align}
where $\sigma(x_1, x_2)$ is the geodesic distance
\begin{align}
	\sigma(x_1, x_2)=2\arcsinh\p{\frac{(r_1-r_2)^2+(\vec{y}_1-\vec{y}_2)^2}{2r_1r_2}}.\label{eq:geodesic}
\end{align}
We use this to define the AdS harmonic functions\footnote{Our normalization agrees with~\cite{Penedones:2007ns},~\cite{Giombi:2021uae} and~\cite{Carmi:2018qzm}, but is different to~\cite{Cuomo:2021kfm}.}
\begin{align}
	\Omega_\nu(X_1, X_2)&=\frac{i\nu}{2\pi}\p{G_{\AdS}^{\frac{d-2}{2}+i\nu}(X_1, X_2)-G_{\AdS}^{\frac{d-2}{2}-i\nu}(X_1, X_2)}.
\end{align}
These are Laplace eigenfunctions satisfying
\begin{align}
	\nabla_{X_1}^2\Omega_\nu(X_1, X_2)=-\p{\frac{(d-2)^2}{4}+\nu^2}\Omega_\nu(X_1, X_2).
\end{align}
with the normalization
\begin{align}
	\Omega_\nu(X, X)&=\frac{\Gamma(\tfrac{d-2}{2})|\Gamma(\tfrac{d-2}{2}+i\nu)|^2}{4\pi^{\tfrac{d}{2}}\Gamma(d-2)|\Gamma(i\nu)|^2}.\label{eq:coincidentOmega}
\end{align}

We can use these functions to give the harmonic decomposition of any function $F(X_1, X_2)$ that is invariant under $\AdS$-isometries,
\begin{align}
	F(X_1, X_2)&=\int_{-\infty}^\infty\,d\nu \tl{F}(\nu)\Omega_\nu(X_1, X_2).
\end{align}
The function $\tl{F}(\nu)$ can then be found by the inverse transform
\begin{align}
	\tl{F}(\nu)=\frac{1}{\Omega_\nu(X, X)}\int d^{d-1}X_1 \sqrt{g(X_1)} F(X_1, X_2)\Omega_\nu(X_1, X_2).\label{eq:inv}
\end{align}

This can be generalized to isometry-invariant functions on $\AdS_{d-1}\times S^1$, by combining it with a Fourier decomposition. For example, it is known~\cite{Penedones:2007ns, Carmi:2018qzm} that the bulk-to-bulk $\AdS_{d-1}$ propagator is given by
\begin{align}
	G_{\AdS}^{\Delta}(X_1, X_2)&=\int_{-\infty}^\infty d\nu\frac{1}{\nu^2+(\Delta-\tfrac{d-2}{2})^2}\Omega_\nu(X_1, X_2).\label{eq:spectral_prop}
\end{align}
The bulk-to-bulk propagator on $\AdS_{d-1}\times S^1$ can be computed as
\begin{align}
	G_{\mathrm{bb}}(x_1, x_2) &=\frac{1}{2\pi}\sum_{s\in\Z+v} e^{is\theta_{21}}G_{\mathrm{bb}}^{\Delta_s}(X_1, X_2)\label{eq:decomp}
\end{align}
where
\begin{align}
	\Delta_s&=\frac{d-2}{2}+\sqrt{\sigma_0^{\mathrm{IR}}+s^2}.
\end{align}
This can be combined with~\eqref{eq:spectral_prop} to obtain that 
\begin{align}
	G_{\mathrm{bb}}(x_1, x_2)&=\sum_{s\in \Z+v} \frac{e^{is\theta_{21}}}{2\pi} \int_{-\infty}^\infty d\nu\frac{1}{\nu^2+(\Delta_s-\tfrac{d-2}{2})^2}\Omega_\nu(X_1, X_2).\label{eq:spectral}
\end{align}
The poles in the spectral representation of $G_{\mathrm{bb}}(x_1, x_2)$ correspond to the scaling dimensions of exchanged defect operators. This motivates us to consider the spectral representation of the $G_{\delta\sigma\delta\sigma}=\braket{\delta\sigma\delta\sigma}$ to determine the dimensions of defect operators exchanged in it.

In order to do so, we consider the spectral representation of the inverse $G^{-1}_{\delta\sigma\delta\sigma}$,
\begin{align}
	G^{-1}_{\delta\sigma\delta\sigma}(x_1, x_2)&=\frac{1}{2\pi}\sum_{\ell\in \Z} e^{i\ell\theta_{21}} \int_{-\infty}^\infty d\nu\,\tl{B}_{\ell}(\nu)\Omega_\nu(X_1, X_2).
\end{align}
To see the relation of $\tl{B}_\ell$ to the spectral representation of $\braket{\delta\sigma(x_1) \,\delta\sigma(x_2)}$, we note that they are the integral kernels of inverse operators. Under convolution,
\begin{align}
	(F*G)(X_1, X_2)=\int d^{d-1}X_3\sqrt{g(X_3)}F(X_1, X_3)G(X_3, X_2)
\end{align}
we have the property~\cite{Costa:2012cb} that 
\begin{align}
	\tl{F*G}(\nu)=\tl{F}(\nu)\tl{G}(\nu),
\end{align}
where $\tl{F}$ and $\tl{G}$ are defined by~\eqref{eq:inv}. This implies that the spectral density of $G_{\de\s\de\s}$ is inverse to that of $G^{-1}_{\de\s\de\s}$ and is therefore given by $\tl{B}_{\ell}(\nu)^{-1}$. Since the scaling dimensions of exchanged operators correspond to poles in the spectral density of $G_{\de\s\de\s}$, we conclude that they are given by the zeros of $\tl{B}_{\ell}(\nu)$. Note that $\ell$ gives the $U_v$ charge of the exchanged operator.

Explicitly, $\tl B_\ell(\nu)$ can be computed as
\begin{align}
	\tl{B}_\ell(\nu)=\frac{1}{\Omega_\nu(X, X)}\int d^dx_1\,\sqrt{g(x_1)}\,e^{-i\ell\theta_{21}} G^{-1}_{\delta\sigma\delta\sigma}(x_1, x_2)\Omega_\nu(X_1, X_2).
\end{align}
Plugging in~\eqref{eq:deltasigmadeltasigma}, we see that we need to evaluate
\begin{align}
	&\frac{1}{\Omega_\nu(X, X)}\int d^dx_1\,\sqrt{g(x_1)}\,e^{-i\ell\theta_{21}} G_{\mathrm{bb}}(x_1, x_2)G_{\mathrm{bb}}(x_2, x_1)\Omega_\nu(X_1, X_2)\\
	=&\frac{2\pi}{\Omega_\nu(X, X)}\sum_{s\in \Z+v}\int d^{d-1}X_1\,\sqrt{G(X_1)}\, G^{\Delta_s}_\AdS(X_1, X_2)G^{\Delta_{s-\ell}}_\AdS(X_2, X_1)\Omega_\nu(X_1, X_2).
\end{align}
Fortunately an explicit formula for this can be found, as
\begin{align}
	&\tilde{C}_{\Delta_1\Delta_2}(\nu)=\frac{1}{\Omega_\nu(X, X)}\int d^{d-1}X_1\,\sqrt{G(X_1)}\, G^{\Delta_1}_\AdS(X_1, X_2)G^{\Delta_2}_\AdS(X_2, X_1)\Omega_\nu(X_1, X_2)\\
	=&\sum_{n=0}^\infty\frac{\Delta_1+\Delta_2-\frac{d}{2}+1+2n}{\nu^2+(\Delta_1+\Delta_2+1-\frac{d}{2}+2n)^2}\\
	&\quad\x\frac{
		(\tfrac{d-2}{2})_n\G(\De_1+n)\G(\De_2+n)(\De_1+\De_2+n-d+2+1)_n\G(\De_1+\De_2-\tfrac{d}{2}+1+n)
	}{
		2\pi^{d/2-1}	n!\G(\De_1-\frac{d}{2}+n)\G(\De_2-\frac{d}{2}+n)\G(\De_1+\De_2+2n)
	}.\label{eq:cTilde}
\end{align}
A derivation for this is given in Appendix~\ref{app:tlC}. Combining this with~\eqref{eq:spectral}, we can find an explicit formula for $\tl{B}_\ell(\nu)$.
\begin{align}
	&\tl{B}_\ell(\nu)=\nn\\
	&\frac{J^2}{8\pi^2}\p{\frac{1}{\nu^2+(\ell+v)^2-v^2+\frac{(d-2)^2}{4}}+\frac{1}{\nu^2+(\ell-v)^2-v^2+\frac{(d-2)^2}{4}}}
	-\frac{1}{4\pi}\sum_{s_1\in \Z+v}\tilde{C}_{\Delta_s\Delta_{s-\ell}}(\nu).
\end{align}

\begin{figure}[t]
	\centering
	\caption{}
	\begin{subfigure}[t]{0.49\textwidth}
		\includegraphics[width=\linewidth]{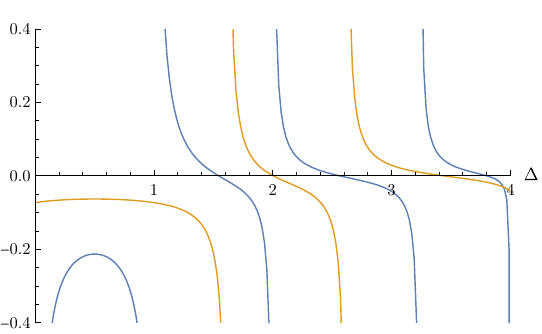}
		\caption{$v=0$}
		\label{fig:S12Bounds}
	\end{subfigure}
	\hfill
	\begin{subfigure}[t]{0.49\textwidth}
		\includegraphics[width=\linewidth]{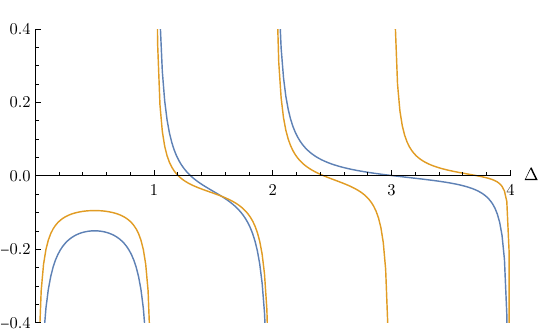}
		\caption{$v=\frac{1}{2}$}
		\label{fig:S12Alphas}
	\end{subfigure}
	\hfill
	\vfill
	\begin{subfigure}[t]{0.49\textwidth}
		\includegraphics[width=\linewidth]{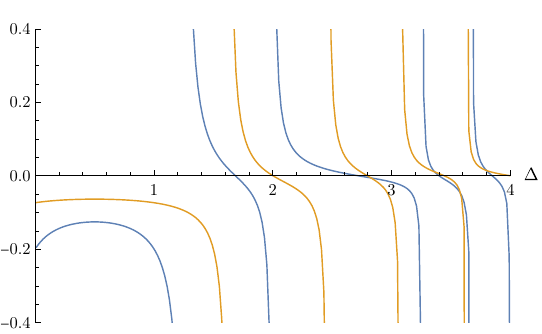}
		\caption{$v=v^*$}
		\label{fig:S12Betas}
	\end{subfigure}
	\caption{Plots of $\tl{B}_0(\frac{1}{2}+\Delta)$ (in blue) and $\tl{B}_1(\frac{1}{2}+\Delta)$ (in orange) in $d=3$ for $v\in\{0, \frac{1}{2}, v^*\}$.}
	\label{fig:defect_dims}
\end{figure}

The zeros of $\tl{B}_\ell(i(\Delta-\frac{d-2}{2}))$ in $\De$ correspond to scaling dimensions of $U_v$-charge $\ell$ operators exchanged in the two-point function of $\delta\sigma$. We can evaluate this function numerically, by truncating the sum after a finite (but large) number of terms. These functions are plotted in figure~\ref{fig:defect_dims} for $v=0, \tfrac{1}{2}, v^*$, and $\ell=0, 1$ in $d=3$.

We extract from these plots scaling dimensions of several defect operators (in $d=3$), which we list to 4 significant figures,
\begin{center}
	\begin{tabular}{ | m{2cm} | m{1.5cm}| m{2.5cm} | m{2.5cm} | m{2.5cm} | } 
		\hline
		$\hat{\mathcal{O}}$ & $\ell_{\hat{\mathcal{O}}}$& $\Delta_{\hat{\mathcal{O}}}$ at $v=0$ & $\Delta_{\hat{\mathcal{O}}}$ at $v=\tfrac{1}{2}$ & $\Delta_{\hat{\mathcal{O}}}$ at $v=v^*$ \\ 
		\hline
		$\Re(\hat{\Psi}^1_v)$ & 0 & 1.543 & 1.309 & 1.000\\ 
		\hline
		$s^-$ & 0 & 2.557 & 2.000 & 1.684 \\ 
		\hline
		$D^i$ & 1 & 2.000 & 2.000 & 2.000 \\ 
		\hline
	\end{tabular}
\end{center}
where $\ell_{\hat{\mathcal{O}}}$ is the $U_v$-charge of the operator. In $v=0$, $d=3$ where this is a pinning field defect, the operators with $U_v$-charge $\ell=0$ have been analysed in~\cite{Cuomo:2021kfm}, and identified with operators in a $4-\varepsilon$ expansion.

The simplest operator to understand is $D^i$ the lightest operator with $U_v$-charge 1, which is the displacement operator. This can be seen as it always has a protected dimension given by $\Delta_D=2$. This provides a good consistency check of our computation. Note that there is no zero in $\tl B_1$ at $\De=2$ in figure~\ref{fig:defect_dims} for $v=\half$. This is because the zero at $\De=2$ collides with a pole precisely at $v=\half$; this implies that the displacement operator decouples from $\<\de \s\de \s\>$ at $v=\half$ at least to the leading order in large $N$. The zero at $\De=2$ is present for all other values of $v\in [0,v^*]$. The decoupling of $D$ is consistent with the Ward identity that relates the two-point function $\<\s D\>$ to the one-point function $\<\s\>$~\cite{Billo:2016cpy} equation~\eqref{eq:saddle} implies that $\<\s\>=0$ in $d=3$ and at $v=\half$.\footnote{We thank the anonymous referee for pointing this out.}

We then note that at $v=0$, the operator we are identifying with $\Re(\hat{\Psi}^1_v)$ has scaling dimension $\Delta_{\Re(\hat{\Psi}^1_v)}=1.543\dots$, which agrees with the dimension found for this operator in~\cite{Cuomo:2021kfm} at $v=0$. Additionally, at $v=v^*$ it agrees with $\Delta_{\Re(\hat{\Psi}^1_v)}=1$ in the pure monodromy defect, as expected. We will also analyse this scaling dimension near $v=v^*$ in conformal perturbation theory in section~\ref{sec:CPT}.

\begin{figure}[t]
	\centering
	\includegraphics[width=0.5\linewidth]{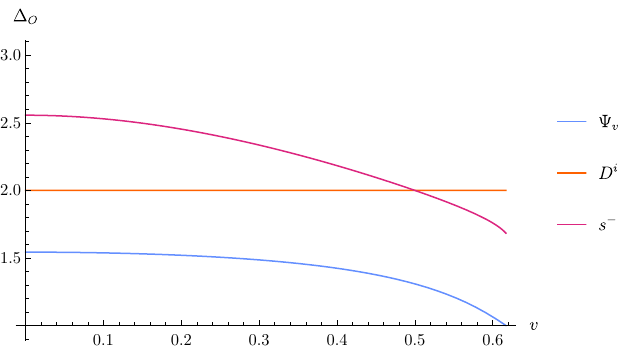}
	\caption{A plot of the scaling dimensions of the defect operators $\Psi_v$, $D^i$, and $s^-$ for $0\leq v\leq v^*$ in $d=3$.}
	\label{fig:OpDims}
\end{figure}

Finally, we denote the next-to leading operator with $\ell$ by $s^-$ since in the case where $v=0$ it coincides with the operator $s^-$ in~\cite{Cuomo:2021kfm}. In particular, our scaling dimension at $v=0$ agrees with~\cite{Cuomo:2021kfm}. Interestingly, from figure~\ref{fig:OpDims} it can be seen that for $v=\frac{1}{2}$ the spectral density $\tl{B}_0(i(\Delta-\frac{d-2}{2}))$ does not have a zero at the location of this operator. By continuity though, we would expect an operator of scaling dimension $\Delta_{s^-}=2.000$ to exist in this theory, but its OPE coefficient is zero at $v=\tfrac{1}{2}$. As the symmetry at $v=\tfrac{1}{2}$ is enhanced from $\mathrm{U}(N-1)$ to $\mathrm{O}(2N-2)$, we conjecture that this is because $s^-$ is an operator which is invariant under $\mathrm{U}(N-1)$, but not under $\mathrm{O}(2N-2)$. Although $\Delta_{s^-}=2.000$ at $v=\tfrac{1}{2}$, we do not think that this scaling dimension is protected, as there seems to be no reason for an operator with this protected scaling dimension and $U_v$-charge to exist. We therefore expect that this is merely a large $N$ artefact, which will receive corrections in a $1/N$-expansion.

\subsection{Conformal perturbation theory in the limit $v\nearrow v^*$}
\label{sec:CPT}

Recall that $v\leq v^*$ is the condition under which the operator $\Re(\hat{\Psi}_v)$ that we are deforming by in~\eqref{eq:action3} is relevant. Therefore, for $v=v^*-\varepsilon$ with small $\varepsilon>0$ this operator is weakly relevant, and the RG flow can be studied through the lens of conformal perturbation theory. In this section we will perform the most basic analysis by computing the leading-order anomalous dimension of $\Re(\hat{\Psi}_v)$ at the IR fixed point. We will then compare the result to the scaling dimension obtained in section~\ref{sec:large_n} through the spectral decomposition of $\<\de\s\de\s\>$.

Let $\delta=d-2-\hat\Delta_v^\text{UV}$, where we have added the superscript UV to stress that we refer to the scaling dimension at the UV monodromy defect fixed point. In the simplest weakly-relevant RG flows, at the leading order in $\de$ the fixed point is found by balancing the linear and quadratic terms in the beta function of the coupling $h$,
\begin{align}
	\beta(h)=-\de h+\beta_2 h^2+\mathcal{O}(h^3).
\end{align}
This $\beta$-function has the UV fixed point at $h=0$ and the IR fixed point at 
\begin{align}
	h^*\overset{?}{=}\frac{\delta}{\beta_2}+O(\de^2),
\end{align}
if $\beta_2\neq0$. An immediate consequence of this analysis is that
\begin{align}
	\beta'(h^*)\overset{?}{=}-\beta'(0)=\delta,
\end{align}
and so that the scaling dimension of $\Re(\hat{\Psi}_v)$ at the IR fixed point would be
\begin{align}
	\hat\Delta_v^\text{IR}=d-2+\beta'(h^*)\overset{?}{=}d-2+\delta+O(\de^2).
\end{align}
However, as we now explain, in our case $\b_2=0$ and this analysis has to be modified.

This is because $\beta_2$ is proportional to the three-point function of $\Re(\hat{\Psi}_v)$, which can be seen to vanish after writing $\Re(\hat{\Psi}_v)=\half (\hat{\Psi}_v+\bar {\hat{\Psi}_v})$ and using the transverse rotation symmetry of the UV fixed point. Therefore, the beta function starts at a higher order,
\begin{align}
	\beta(h)=-\de h+\beta_3 h^3+\dots.
\end{align}
The coefficient $\b_3$ is determined by the four-point function of $\Re(\hat{\Psi}_v)$, which is not forced to be zero by any symmetry. We therefore expect $\b_3\neq 0$.

The roots of $\beta$ are then $h=0$ and $h=\pm h^*$ where
\begin{align}
	h^*=\sqrt{\frac{\delta}{\beta_3}}+O(\de^{3/2}).
\end{align}
Furthermore, we find
\begin{align}
	\beta'(h^*)=\beta'(-h^*)=2\delta,
\end{align}
which implies that at the IR fixed points
\begin{align}
	\Delta^\text{IR}_{v}=d-2+2\delta.
\end{align}

We can compare this prediction to the large-$N$ results of section~\ref{sec:large_n}. We find that numerically to 4 significant figures,
\begin{align}
	\lim_{v\nearrow v^{*}} \frac{\hat{\Delta}{}_{v}^\text{IR}-(d-2)}{\hat{\Delta}{}_{}^\text{UV}-(d-2)}\approx -2.000,
\end{align}
as expected, in agreement with the prediction of conformal perturbation theory. Note the denominator is obtained using~\eqref{eq:dimensions}, valid in the UV monodromy DCFT, while the numerator is determined from the leading zero of $\tl B_0$.

Note that having a fixed point with a real value of $h$ for $\de>0$ (which is what seems to happen) requires $\b_3>0$. Therefore, for $\de=0$ the perturbation by $h$ is marginally irrelevant. This means that at $\de=0$, there is no relevant flow triggered by $h$. On the other hand, the RG flows that preserve the same symmetry as the would-be monodromy pinning defect and terminate in the IR at the monodromy defect, will have logarithmic corrections to scaling, coming from the beta-function for $h$.

\section{$4-\varepsilon$ expansion}
\label{sec:epsilon}

In this section we perform the leading-order analysis in $4-\varepsilon$ expansion. To do so, we can begin with the flat-space action:
\begin{align}
	S=\int d^dx \p{\frac{1}{2}\partial_\mu\bar\Phi{}^I(x)\partial^\mu\Phi^I(x)+\frac{\lambda}{4}\p{\bar\Phi{}^I(x)\Phi^I(x)}^2}+h\int_{\mathcal{D}}d^{d-2}\vec{y}\,
	\Re(\hat{\Psi}^1_v(\vec{y})).
\end{align}
We note that we are now working at finite $N$.

An immediate issue with the above action is that the operator $\Re(\hat{\Psi}^1_v(\vec{y}))$ on the UV monodromy defect has dimension $1+v+O(\varepsilon)$~\cite{Giombi:2021uae}, and therefore for generic $v\in [0,1)$ it is strongly relevant. However, the bulk is weakly-coupled and we expect that the RG flow can still be analysed to the leading order by solving the bulk equations of motion with the source term provided by $\Re(\hat{\Psi}^1_v(\vec{y}))$.

Away from the defect, the equation of motion for this theory is
\begin{align}
	-\partial^2\Phi^I(x)+\lambda\bar\Phi{}^J\Phi^J\Phi^I=0.
\end{align}
In the cylindrical coordinates this becomes
\begin{align}
	(\partial_y^2+\partial_r^2+r^{-1}\partial_r+r^{-2}\partial_\theta^2)\Phi^I=\lambda\bar\Phi{}^J\Phi^J\Phi^I.\label{eq:EoM}
\end{align}

The $\mathrm{O}(2N-2)$ symmetry requires that $\braket{\Phi^I(x)}=0$ if $I\neq1$. Additionally, conformal symmetry along the defect and the $\mathrm{U}(1)$ symmetry generated by $Q_s$~\eqref{eq:charge} constrain the form of $\braket{\Phi^I(x)}$ to be
\begin{align}
	\braket{\Phi^I(x)}=\delta^{I1}e^{iv\theta}\varphi(r),
\end{align}
where $\varphi(r)$ is a function only depending on $r$. It is constrained by the equation of motion~\eqref{eq:EoM} to obey
\begin{align}
	\varphi''(r)+r^{-1}\varphi'(r)-v^2r^{-2} \varphi(r)=\lambda\bar\varphi(r)\varphi(r)^2\label{eq:EoMVarphi}
\end{align}
away from the defect.

It is possible to solve this equation of motion with a source term to obtain a non-trivial profile $\vf(r)$ for all values of $0<r<+\oo$. This solution would track the entire RG flow from the UV to the IR of the theory. However, as we are only interested in the IR fixed point, it is much easier to search for a scale-invariant solution at large $r$, which has to take the form
\begin{align}
	\varphi(r)=c_v r^{-1},
\end{align}
based upon the leading-order dimension of $\Phi$, $\De_\Phi=1+O(\varepsilon)$. The only way for this to be consistent with~\eqref{eq:EoMVarphi} is if
\begin{align}
	(1-v^2)=\l|c_v|^2.
\end{align}

The critical value of the bulk coupling at the critical point is given by~\cite{kleinert2001critical}
\begin{align}
	\l^*=\frac{8\pi^2\varepsilon}{N+4}+\mathcal{O}(\varepsilon^2)
\end{align}
which shows that to leading order in $\varepsilon$,
\begin{align}
	|c_v|^2=\frac{(1-v^2)(N+4)}{8\pi^2\varepsilon}.
\end{align}
In other words,
\begin{align}\label{eq:4minuseps}
	\<\Phi^1(x)\> = e^{iv(\theta-\theta_0)} \sqrt{\frac{(1-v^2)(N+4)}{8\pi^2\varepsilon}}\frac{1}{r},
\end{align}
where $\theta_0$ is determined by the phase of the UV coupling through the RG flow. As in section~\ref{sec:large_n}, we set $\theta_0=0$ by a transverse rotation.

This can be compared to our large-$N$ analysis in section~\ref{sec:large_n}. We determine the value of $J$ in $d=4-\varepsilon$ in appendix~\ref{app:Gbb}, and using~\eqref{eq:largeN1pt} find that to leading order in $\varepsilon$ and $1/N$,
\begin{align}
		\braket{\Phi^1(x)}=e^{iv\theta}\sqrt{\frac{(1-v^2)N}{8\pi^2\varepsilon}}\frac{1}{r}\label{eq:4-epsilon_1pt}.
\end{align}
This agrees with~\eqref{eq:4minuseps} to the leading order at large $N$.

\section{Conclusions}
\label{sec:conclusions}
In this work, we have explored a spinning conformal defect in the $\mathrm{O}(2N)$ model. Specifically we looked at the IR fixed point of an RG flow triggered by a relevant defect operator on a monodromy defect in the $\mathrm{O}(2N)$ model. We examined this spinning DCFT in large-$N$ and $4-\varepsilon$ expansions.

The main results in the large-$N$ expansion include the leading-order scaling dimensions of various defect operators, and the value of bulk one-point function $\braket{\Phi^I}$. These results were then compared to corresponding results to leading order in the $4-\varepsilon$-expansion, and also to expectations from conformal perturbation theory for $v\approx v^*$. Agreement between these results provides a useful cross-check for the validity of our results, as does comparing them in the case $d=3$, $v=0$ to the results of~\cite{Cuomo:2021kfm}.

There are several natural directions in which this work could be extended. Firstly, it would be interesting to study the pinning monodromy DCFT in more detail. For instance, developing a systematic $4-\varepsilon$ expansion seems like a conceptually important task. In particular, one can try comparing, in the spirit of~\cite{deSabbata:2024xwn}, the results of this expansion at $v=0$ with the more traditional $4-\varepsilon$ expansion of the pinning field defect in $d=3$. The conformal perturbation theory around $v\approx v^*$ can also be explored further.

Secondly, we considered only one possible relevant perturbation and it would be interesting to explore others, such as simultaneous perturbations by multiple relevant operators. It would also be interesting to find other examples of spinning DCFTs and to investigate their Weyl anomalies~\cite{Kravchuk:2024qoh} as well as integrated identities involving tilt and displacement operators~\cite{Gabai:2025hwf,Gabai:2025zcs,Girault:2025kzt, Drukker:2022pxk, Kong:2025sbk,DKKFuture}.

\section*{Acknowledgments}

We would like to thank Gabriel Cuomo, Nadav Drukker, Christopher Herzog, and Andreas Stergiou for helpful insights and discussions about the topic of this paper. AR would also like to thank
King’s College London for providing funding while this research was carried out. The work of PK was funded by UK Research and Innovation (UKRI) under the UK government's Horizon Europe funding Guarantee [grant number EP/X042618/1] and the Science
and Technology Facilities Council [grant number ST/X000753/1].

\appendix

\section{Bulk-to-defect OPE of a monodromy defect in the $\mathrm{O}(2N)$ model}
\label{app:OPE}
The bulk-to-defect OPE in a monodromy defect can be written as~\cite{Giombi:2021uae}
\begin{align}
	\Phi^I(x)=\sum_{s\in\Z+v} C_s \frac{e^{is\theta}}{r^{\Delta_\Phi-\hat\Delta_s}}\sum_{m=0}^\infty \frac{(-1)^m r^{2m}(\vec\partial_y^2)^m}{m!\,2^{2m} (\hat\Delta_s+2-\frac{d}{2})_m}\hat{\Psi}^I_s(\vec{y}).\label{eq:OPE}
\end{align}
and
\begin{align}
	\bar\Phi{}^I(x)=\sum_{s\in\Z+v} C_s^*\frac{e^{-is\theta}}{r^{\Delta_\Phi-\hat\Delta_s}}\sum_{m=0}^\infty \frac{(-1)^m r^{2m}(\vec\partial_y^2)^m}{m!\,2^{2m} (\hat\Delta_s+2-\frac{d}{2})_m}\hat{\bar{\Psi}}{}^I_s(\vec{y}).\label{eq:OPEbar}
\end{align}

For $v=0$ or $v=\tfrac{1}{2}$, this can be written in terms of the real fields as
\begin{align}
	\phi^a(x)&=\sum_{s\in\Z+v} \frac{e^{is\theta}}{r^{\Delta_\Phi-\hat\Delta_s}}\sum_{m=0}^\infty \frac{(-1)^m r^{2m}(\vec\partial_y^2)^m}{m!\,2^{2m} (\hat\Delta_s+2-\frac{d}{2})_m}\hat\chi^a_s(\vec{y}),
\end{align}
where
\begin{align}
	\hat\chi^{2I-1}_s(\vec{y})=\frac{\hat{\bar{\Psi}}{}^I_s(\vec{y})+\hat{\Psi}^I_s(\vec{y})}{2},\qquad\hat\chi^{2I}_s(\vec{y})=\frac{i(\hat{\bar{\Psi}}{}^I_{s}(\vec{y})-\hat{\Psi}^I_s(\vec{y}))}{2}.
\end{align}
These obey the reality conditions that $\bar{\hat\chi^a_s}=\hat\chi^a_{-s}$.

\section{Regularization of $G_{\mathrm{bb}}(x, x)$}
\label{app:Gbb}
In this appendix, we shall explain how we regularize $G_{\mathrm{bb}}(x, x)$, as defined in~\eqref{eq:Gbb}. We can use~\eqref{eq:spectral_prop} and~\eqref{eq:coincidentOmega} to find that
\begin{align}
	G_{\AdS}^{\Delta_s}(X, X)&=\int_{-\infty}^\infty d\nu\frac{1}{\nu^2+(\Delta-\tfrac{d-2}{2})^2}\Omega_\nu(X_1, X_2)\\
	&=\frac{-\pi ^{1-\frac{d}{2}} \Gamma \left(\frac{d-2}{2}\right)\Gamma \left(\frac{d-2}{2}+\sqrt{s^2-v^2+\frac{(d-2)^2}{4}}\right)}{4 \cos \left(\frac{\pi  d}{2}\right) \Gamma (d-2) \Gamma \left(-\frac{d-4}{2}+\sqrt{s^2-v^2+\frac{(d-2)^2}{4}}\right)},
\end{align}
for $d>2$, where the integral can be performed by completing the contour in the upper half plane. In the limit $\ell\to\pm\infty$ we find the following asymptotic
\begin{align}
	G_{\AdS}^{\Delta_{\ell+v}}(X, X)&=  \frac{-\pi ^{1-\frac{d}{2}} \sec \left(\frac{\pi  d}{2}\right) \Gamma \left(\frac{d-2}{2}\right)}{4 \Gamma (d-2)}(|\ell|^{d-3}\pm(d-3) v|\ell|^{d-4})+\mathcal{O}(|\ell|^{d-5}).
\end{align}
We can hence see that the sum $G_{\mathrm{bb}}(x, x)=\frac{1}{2\pi}\sum_{s\in \Z+v} G_{\AdS}^{\Delta_s}(X, X)$ will not converge for $d>2$. Therefore, for $2<d<4$, we regularize the sum by explicitly resumming the leading asymptotic term for $d<2$ and analytically continuing,
\begin{align}
	G_{\mathrm{bb}}(x, x)=\frac{1}{2\pi}&\Bigg(G_{\AdS}^{\Delta_v}(X, X)-\frac{\pi ^{1-\frac{d}{2}} \sec \left(\frac{\pi  d}{2}\right) \Gamma \left(\frac{d-2}{2}\right)}{2\,\Gamma (d-2)}\zeta(3-d)\label{eq:coincident}\\
	&+\sum_{\ell=1}^\infty  \p{G_{\AdS}^{\Delta_{\ell+v}}(X, X)+G_{\AdS}^{\Delta_{-\ell+v}}(X, X)+\frac{\pi ^{1-\frac{d}{2}} \sec \left(\frac{\pi  d}{2}\right) \Gamma \left(\frac{d-2}{2}\right)}{2 \Gamma (d-2)}\ell^{d-3}}\Bigg).\nn
\end{align}

Equation \eqref{eq:coincident} is still singular at $d=3$, but can be generalized to this case by continuity in $d$. This leads to $G_{\mathrm{bb}}(x, x)$ being well-defined for any $2<d<4$. We do not know a closed form for the sum, but it can be numerically approximated by truncation in $\ell$.

We also require $G_{\mathrm{bb}}(x, x)$ in $d=4-\varepsilon$ expansion. We find that there is divergence as $\varepsilon\to0$. This divergence can be extracted from the zeta function, and we find that to the leading order in $\varepsilon$,
\begin{align}
	G_{\mathrm{bb}}(x, x)=\frac{v^2-1}{8 \pi ^2 \varepsilon}+O(\varepsilon^0).
\end{align}

\section{An explicit formula for $\tl{C}_{\Delta_1\Delta_2}(\nu)$}
\label{app:tlC}
In this appendix we derive the expression for
\begin{align}\label{eq:Ctarget}
	\tl{C}_{\Delta_1\Delta_2}(\nu)&=\int d^{d-1}X_1\,\sqrt{G(X_1)}\, G_\AdS^{\Delta_1}(X_1, X_2)G_\AdS^{\Delta_2}(X_2, X_1)\Omega_\nu(X_1, X_2),
\end{align}
given in equation~\eqref{eq:cTilde}. In the case $\De_1=\De_2$ this has been computed in~\cite{Carmi:2018qzm}. We follow a similar strategy in that we compute the residues of the poles of $\tl{C}_{\De_1\De_2}(\nu)$ in $\nu$ and then recover $\tl{C}_{\De_1\De_2}(\nu)$ using Mittag-Leffler's theorem.

By AdS-isometry invariance, the integral~\eqref{eq:Ctarget} can be written in terms of the geodesic distance (as defined in \eqref{eq:geodesic}) which gives
\begin{align}
	\tl{C}_{\Delta_1\Delta_2}(\nu)&=\int_0^\infty d\sigma (\sinh\sigma)^{d-2} G_\AdS^{\Delta_1}(\sigma)G_\AdS^{\Delta_2}(\sigma)\Omega_\nu(\sigma).
\end{align}
An explicit expression for $G_\AdS^{\Delta}$ was given in \eqref{eq:propagator}, and  $\Omega_\nu(\sigma)$ is given in~\cite{Carmi:2018qzm} as
\begin{align}
	\Omega_\nu(\sigma)=\frac{\Gamma \left(\frac{d-2}{2}\right) \Gamma \left(\frac{d-2}{2}+i \nu \right) \Gamma \left(\frac{d-2}{2}-i \nu \right)}{4 \pi ^{\frac{d}{2}} \Gamma (d-2) \Gamma (i \nu ) \Gamma (-i \nu)}{}_2F_1\left(\frac{d-2}{2}+i \nu ,\frac{d-2}{2}-i \nu ;\frac{d-1}{2};-\sinh ^2\left(\frac{\sigma }{2}\right)\right),
\end{align}
\begin{align}
	&G_\AdS^{\Delta}(\sigma)=\frac{\Gamma(\Delta)}
	{2\pi^{\tfrac{d-2}{2}}\,\Gamma\!\left(\Delta-\frac{d}{2}+2\right)\big(2\cosh(\sigma)\big)^\Delta}
{}_2F_1\!\left(\frac{\Delta}{2},\frac{\Delta+1}{2};
	\;\Delta-\frac{d}{2}+2;
	\;\frac{1}{\cosh^2(\sigma)}\right).
\end{align}

To simplify the integral, we shall strip off some $\sigma$-independent prefactors, and evaluate,
\begin{align}
	\int_0^\infty d\sigma (\sinh\sigma)^{d-2} g^{\Delta_1}(\sigma )g^{\Delta_2}(\sigma )\omega_\nu(\sigma).
\end{align}
where
\begin{align}
	\omega_\nu(\sigma)&={}_2F_1\left(\frac{d-2}{2}+i \nu ,\frac{d-2}{2}-i \nu ;\frac{d-1}{2};-\sinh ^2\left(\frac{\sigma }{2}\right)\right)
\end{align}
and
\begin{align}
	g^\Delta(\sigma)=\frac{\cosh(\sigma)^{-\Delta}}{\Gamma(-\frac{d}{2}+\Delta +2)} \, _2F_1\left(\frac{\Delta }{2},\frac{\Delta +1}{2};-\frac{d}{2}+\Delta +2;\frac{1}{\cosh ^2\left(\sigma \right)}\right)
\end{align}
It is useful to write $\omega_\nu=\omega^R(\nu)+\omega^R(-\nu)$ where
\begin{align}
	\omega^R(\nu)=\frac{2^{d-3}\G(\tfrac{d-1}{2})\G(-i\nu)}{\sqrt\pi \G(\tfrac{d-2}{2}-i\nu)}e^{-(\frac{d-2}{2}+i\nu)\s}{}_2F_1(\tfrac{d-2}{2},\tfrac{d-2}{2}+i\nu;i\nu+1;e^{-2\s}),
\end{align}
in order to explicitly exhibit large-$\s$ asymptotics.

We hence consider the integral
\begin{align}
	R(\nu)&=\int_0^\infty d\sigma (\sinh\sigma)^{d-2} g^{\Delta_1}(\sigma )g^{\Delta_2}(\sigma )\omega^R_\nu(\sigma).
\end{align}
In order to evaluate it, we introduce the function
\begin{align}
	h^{\De_1 \De_2}(\nu, z)&=(1-z)^{d-2}
{}_2F_1(\De_1,\tfrac{d-2}{2};\De_1-\tfrac{d}{2}+2;z){}_2F_1(\De_2,\tfrac{d-2}{2};\De_2-\tfrac{d}{2}+2;z)\\
&\qquad\qquad\qquad\qquad\times{}_2F_1(\tfrac{d-2}{2},\tfrac{d-2}{2}-i\nu;1-i\nu;z),
\end{align}
so that we can write
\begin{align}\label{eq:intR}
	R(\nu)=\frac{2^{\De_1+\De_2}\G(\tfrac{d-1}{2})\G(i\nu)}{2\sqrt\pi \G(\tfrac{d-2}{2}+i\nu)\G(\De_1-\tfrac{d}{2}+2)\G(\De_2-\tfrac{d}{2}+2)}\int_0^\oo d\s e^{i\nu\s-(\De_1+\De_2-\frac{d-2}{2})\s}h^{\Delta_1\Delta_2}(\nu, e^{-2\s}).
\end{align}

We can expand
\begin{align}\label{eq:taylor}
	h^{\Delta_1\Delta_2}(\nu, z)=\sum_{n=0}^\infty h^{\Delta_1\Delta_2}_n(\nu) z^n.
\end{align}
One can check that the only poles of $\tl{C}_{\Delta_1\Delta_2}(\nu)$ come from the large-$\s$ region in~\eqref{eq:intR}, and their locations and residues are given by
\be
\int_0^\oo d\s e^{i\nu\s-(\De_1+\De_2-\frac{d-2}{2})\s}h(\nu, e^{-2\s})\sim \frac{h^{\Delta_1\Delta_2}_n(-i(\De_1+\De_2-\tfrac{d}{2}+1+2n))}{-i\nu+(\De_1+\De_2-\frac{d}{2}+1+2n)}.
\ee
By evaluating the Taylor series~\eqref{eq:taylor} explicitly, we were able to guess the general form
\be
&h^{\Delta_1\Delta_2}_n(-i(\De_1+\De_2-\tfrac{d}{2}+1+2n))\nn\\
&=
\frac{
	(\tfrac{d-2}{2})_n(\De_1)_n(\De_2)_n(\De_1+\De_2+n-d+3)_n
}{
	n!(\De_1-\frac{d}{2}+1)_n(\De_2-\frac{d}{2}+2)_n(\De_1+\De_2+n-\frac{d}{2}+1)_n\label{eq:guess}
}.
\ee
We have not derived this expression rigorously, but we have verified it to a large order $n$.\footnote{We would like to thank Kamran Salehi Vaziri for pointing out that this result can also be derived from equation~(3.14) of \cite{Fitzpatrick:2011hu} as was done in \cite{Loparco:2023rug}, which proves that expression \eqref{eq:guess} is indeed correct.}

Putting all of this together and using Mittag-Leffler's theorem as in~\cite{Carmi:2018qzm}, we obtain equation~\eqref{eq:cTilde}. Our result agrees with equation~(4.26) in~\cite{Carmi:2018qzm} in the case that $\Delta_1=\Delta_2$.

\newpage
\bibliographystyle{JHEP}
\bibliography{refs}

\end{document}